# A Robust nitridation technique for fabrication of disordered superconducting TiN thin films featuring phase slip events

*Sachin Yadav[1, 2], Vinay Kaushik[3], M. P. Saravanan[3], R. P. Aloysius[1, 2], V. Ganesan[3] and Sangeeta Sahoo[1, 2,]**

[1]*Academy of Scientific and Innovative Research (AcSIR), AcSIR Headquarters CSIR-HRDC Campus, Ghaziabad, Uttar Pradesh, 201002, India.*

[2]*Electrical & Electronics Metrology Division, National Physical Laboratory, Council of Scientific and Industrial Research, Dr. K. S Krishnan Road, New Delhi-110012, India.*

[3]*Low Temperature Laboratory, UGC-DAE Consortium for Scientific Research, University Campus, Khandwa Road, Indore- 452001, India*

**Correspondences should be addressed to S. S. (Email: sahoos@nplindia.org)*

**Abstract**

Disorder induced phase slip (PS) events appearing in the current voltage characteristics (IVCs) are reported for two-dimensional TiN thin films produced by a robust substrate mediated nitridation technique. Here, high temperature annealing of Ti/$Si_3N_4$ based metal/substrate assembly is the key to produce majority phase TiN accompanied by $TiSi_2$ & elemental Si as minority phases. The method itself introduces different level of disorder intrinsically by tuning the amount of the non-superconducting minority phases that are controlled by annealing temperature ($T_a$) and the film thickness. The superconducting critical temperature ($T_c$) strongly depends on $T_a$ and the maximum $T_c$ obtained from the demonstrated technique is about 4.8 K for the thickness range ~12 nm and above. Besides, the dynamics of IVCs get modulated by the appearance of intermediated resistive steps for decreased $T_a$ and the steps get more prominent for reduced thickness. Further, the deviation in the temperature dependent critical current ($I_c$) from the Ginzburg-Landau theoretical limit varies strongly with the thickness. Finally, the $T_c$, intermediate resistive steps in the IVCs and the depairing current are observed to alter in a similar fashion with $T_a$ and the thickness indicating the robustness of the synthesis process to fabricate disordered nitride-based superconductor.

## Introduction

Disordered superconductors have exhibited various fundamental quantum phenomena like superconductor-insulator transition (SIT)[1,2], quantum criticality[3,4], quantum phase slip (QPS)[5] etc. Of particular interest, disordered superconductors are suitable for phase slip (PS) study[5] which may lead to interesting applications in single photon detection and quantum current metrology[6]. The PS mechanism, which relates to a change in time for superconducting order parameter due to phase fluctuations, plays a crucial role for the survival of superconductivity in reduced dimension[7]. Further, the physical mechanism of dissipation in superconductors is associated with PS process[8]. For example, in 1D superconducting nanowire, the phase of the order parameter fluctuates continuously and the phase fluctuations lead to vanishing order parameter momentarily when the phase slips by 2π and thus making the wire resistive at the phase slip centers (PSCs). This is known as PS phenomenon which can be initiated by applied current for a current carrying superconductor[9].

Analogous to PSCs in 1D, phase fluctuations lead to phase slip lines (PSLs) in wide 2D superconducting films that exhibit resistive steps in current-voltage characteristics (IVCs)[10]. Experimentally PSs can be triggered by magnetic doping[11,12], embedding artificial pinning centers[13] or by using magnetic field[7]. Here, we address the effects of disorder on the dynamics of IVCs for quasi 2D TiN thin films. In recent years, ultrathin TiN films have been reported to exhibit SIT[14], quantum criticality[14] and Berezinskii-Kosterlitz-Thouless (BKT) transition[15]. Further, it has been reported that disordered TiN can be a promising candidate for QPS study[16,17]. Here, we report the formation of PSLs by observation of staircase-like structures appearing in the IVCs for disordered TiN thin films. We demonstrate that by tuning the disorder the superconducting-to-normal state transition in the measured IVCs changes from a single step transition to a transition embedded with multiple steps featuring the formation of PSLs. The tuning of disorder level in TiN films has been achieved by adopting a novel substrate mediated nitridation technique[18-20], which employs high temperature annealing of Ti/$Si_3N_4$ based metal/substrate assembly to prepare majority phase TiN thin films accompanied by $TiSi_2$ and elemental Si as minority phases[18]. Conventionally, TiN films are grown either by using reactive dc sputtering of Ti in presence of Ar/$N_2$ gas mixtures[21-26] or by atomic layer deposition[27,28] in presence of $N_2$ gas etc.[29-31] However, a fine tuning of nitrogen concentration is needed to achieve the stoichiometric highly crystalline fcc TiN known to possess higher superconducting critical temperature ($T_c$).

By using this substrate mediated nitridation technique, we have produced 3-20 nm thick TiN films with $T_c$ ranging from 3.0 K to 4.8 K, respectively, and the results are comparable to the best reported values[21,27,29]. Here, we address the effects of $T_a$ and film thickness on the superconducting properties of TiN thin films. The maximum $T_c$ obtained is around 4.8 K for the film with thickness 20 nm and annealed at 820°C. $T_c$ decreases with the decrease in $T_a$ & film thickness. Further, the formation of PSLs and their evolution in the IVCs have been studied under the application of external fields. Finally, we have investigated the effects of $T_a$ and film thickness on the temperature dependent critical current ($I_c$) by using Ginzburg-Landau (GL) theory[32] and we have found that the $I_c$ and the PSLs vary in a similar fashion with $T_a$ and the film thickness.

## Results and discussion

The temperature dependent resistance [$R(T)$] measurements on thin films-based samples are presented in Fig. 1. The device configuration in conventional four probe measurement scheme for a representative sample is shown in the inset of Fig. 1(a). The width of the film is around 500 μm and the distance between the two voltage leads is around 1.1 mm. A set of zero field $R(T)$, measured on the batch of 12 nm thick samples annealed at different $T_a$, is presented in Fig. 1(a). During the course of $R(T)$ measurements from room temperature to 2 K, the samples undergo normal (NM) to superconducting (SC) phase transition which strongly depends on $T_a$. It is evident from Fig. 1(a) that the NM-SC transition shifts towards lower temperature for lower values of $T_a$. For example, the highest $T_c$ appears for $T_a$ = 820 °C, whereas, we do not observe any transition down to 2 K for the sample annealed at 650 °C. The various critical temperatures are defined in Fig. 1(b) which displays the thickness dependent $R(T)$ data for samples annealed at 820 °C. Here, the film thickness varies from 20 nm to 3 nm. The $T_c$ is defined as the temperature at which the derivative $dR/dT$ becomes maximum, which is shown by the dashed vertical line in Fig. 1(b) for the representative sample of thickness 3 nm. The maximum $T_c$ of about 4.8 K is obtained for the sample TN1 of 20 nm thickness. With thickness decreasing from 20 nm to 12 nm as that for sample TN2, the $T_c$ is reduced to 4.76 K indicating a very little change or saturation in $T_c$ for the films with thickness ≥ 12 nm. With further reduction in the film thickness, $T_c$ reduces from that of the maximum $T_c$ of TN1 (~20 nm) by 0.4 K, 1.2 K and 1.8 K for samples TN3 (~8 nm), TN4 (~4 nm) and TN5 (~3 nm), respectively. This implies that the effect of film thickness on $T_c$ becomes prominent for the thickness around 10 nm and below. The reduction in film thickness increases the normal state resistance ($R_n$). Intially, the change in $R_n$ from 20 nm (27.6 Ω) to 12 nm (86.8 Ω) is more than 3 times. With further decreasing thickness, $R_n$ increases to more than 5.5 times for 8 nm (154 Ω), 11.8 times for 4 nm (327 Ω) and 27.8 times for 3 nm (769 Ω) with respect to 20 nm thick sample. However, the change in $R_n$ with respect to $T_a$, as shown in Fig. 1(a), is apparently random and will be discussed later.

Further, we have collected the $T_c^{Onset}$ and $T_c^{Zero}$ values for samples with various thickness for $T_a$ equal to 820 °C, 780 °C and 750 °C and those are displayed with respect to thickness in Fig. 1(c). For $T_a$ = 820 °C and 780 °C, $T_c$ strongly depends on thickness in the lower thickness regime (thickness ~ 4 nm and lower), whereas, for $T_a$ = 750 °C, the variation in the $T_c$ with thickness is very little and the same is marked by the dotted circle. The overall variation of the $T_c$ on $T_a$ and film thickness is displayed by color plot in Fig. 1(d) which is divided into several zones. Based on the increasing order of the $T_c$ , the regimes I-to-IV are separated from each other. Regime-I, the lowest $T_c$ regime, corresponds to a very little change of about 0.1 K in $T_c$ with thickness in the range of 4 nm and below for $T_a$ < 780°C. In comparison with regime-I, regime-II is noticeably different. Here, for $T_a$ > 780°C, $T_c$ remains almost unchanged for a very narrow thickness range, whereas, for $T_a$ < 780°C, a wide span of thickness delivers almost the same $T_c$. Hence in regime-II, the $T_c$ shows strong dependence on thickness for higher $T_a$. In regime-III, the efffect of increased $T_a$ or increased thickness results in same direction towards increasing $T_c$. Finally, the highest $T_c$ zone regime-IV is originated mainly from higher $T_a$ (≥ 780°C) and here, the $T_c$ is observed to saturate for increased thickness. However, for the case of higher film thickness (~ 4 nm and above), the $T_c$ strongly varies with $T_a$ from 3.42 K to 4.76 K. Therefore, $T_a$ plays a significant role in changing the $T_c$ ranging from 4.76 K for 820 °C to no transition down to 2 K for 650 °C for 12 nm thick films. Apart from the $T_c$, the influence of $T_a$ is also observed on $R_n$ as already evident in Fig. 1(a). Here, $R_n$ reduces with decreasing $T_a$ from 820 °C to 750 °C and further reduction in $T_a$ leads to increased $R_n$. Unlike $T_c$, the dependence of $R_n$ on $T_a$ is not straightforward and that is mainly due to the presence of minority phases like $TiSi_2$ and Si that are generated and modulated by the annealing process[18]. Here, $R_n$ varies in the range between 60-87 Ω for $T_a$ ranging from 820 °C to 700 °C with its maximum value of about 87 Ω occuring at $T_a$ = 820 °C. In order to have a clear understanding of the anomaly in $R_n$ variation with $T_a$, we have displayed X-ray photoelectron spectroscopy (XPS) core-level binding energy spectra for Ti 2p, N 1s and Si 2p from the reference samples in Fig. 2(a), (b) and (c), respectively. These reference samples were fabricated simultaneously with the batch of samples shown in the Fig. 1(a). Ti 2p and N 1s spectra clearly indicate the presence of TiN as the dominant phase for almost all the studied $T_a$, whereas, there is a striking difference in Si 2p spectra presented in Fig. 2(c) which indicates a clear shifting from the $TiSi_2$ phase towards elemental Si for $T_a$ > 780 °C. As the resistivity of $TiSi_2$[33] is lower than that of TiN[34] and of Si, the reduced $R_n$ for $T_a \leq$ 780 °C compared to that of $T_a$ = 820 °C can be originated from the amount of $TiSi_2$ present in the composite film. Further, the relative concentration of minority $TiSi_2$ phase changes with $T_a$[18] and hence, $R_n$ is expected to change with $T_a$. As evident from Fig. 2(c) at $T_a$ = 820 °C, a majority of the decomposed Si from $Si_3N_4$ substrate remains as elemental Si which in turn contribute to the increased resistance in the normal state[18]. However, the change in $R_n$ with respect to thickness [Fig. 1(b)] is much more than the change due to the variations in $T_a$ [Fig. 1(a)]. For comparison, we have presented $R_n$ with respect to $T_a$ and thickness by using 3-D color plots in Fig. 2(d). It is apparent that $R_n$ is mainly dominated by the thickness variations.

Finally, from Fig. 1&2, we have demonstrated the variation of $T_c$ & $R_n$ with thickness & $T_a$ for disordered TiN thin films accompanied by nonsuperconducting minority phases of $TiSi_2$ and Si[18]. As $T_a$ plays a huge role in the decomposition of $Si_3N_4$ and diffusion of Si & N elements inside the film[18], the influential role of $T_a$ on particularly the superconducting properties of TiN thin films is also confirmed by the present study. It is noteworthy to mention that the achieved best $T_c$ by employing the current synthesis technique can be comparable, and some cases can be even better than, the reported conventional methods [Table S1 in the Supplementary Information (SI)].

So far, we have discussed the influence of film thickness & $T_a$ on $T_c$ and $R_n$. Now, we consider the effect of film thickness and $T_a$ on the IVCs for samples prepared under different growth conditions. In Fig. 3, we display IVCs of three selective samples, namely, TN4, TN8 & TN12, that are annealed at 820 °C, 780 °C & 750 °C, respectively. Each of these selected samples are of 4 nm thickness which is less than their respective coherence length $\xi(0)$ [Fig. S5 in SI]. This indicates that the samples are in the 2D limit, where phase fluctuations lead to PSLs[35] and *IVCs* exhibit resistive steps[20,36,37]. In Fig. 3(a), we display a set of zero field *R(T)* measured on the aforementioned three representative samples. Similar to Fig. 1(a), here also we observe a shifting of the NM-SC phase transition towards lower temperature for reduced $T_a$. For example, $T_c$ changes from 3.6 K for sample TN4 (~820 °C) to 3.12 K for sample TN8 (~780 °C) to 2.5 K for sample TN12 (~750 °C). However, very little variation in $R_n$ is observed among the samples. First, we present the IVC isotherms for sample TN4 in Fig. 3(b) which shows a single step sharp switching during the phase transition from the superconducting-to-metallic state at temperature far below the $T_c$. The current at which the switching occurs is known as the critical current ($I_c$), whereas, the current corresponding to the onset of superconducting state from the normal state during the downward sweeping direction is defined as the retrapping current ($I_r$)[Fig. S6 in SI]. For all the measured samples, these two characteristic currents are different and the IVCs appear to be hysteretic with respect to the current sweeping direction. With increasing temperature, the hysteresis reduces and finally vanishes at $T_c$ as shown in Fig. S6 in SI. Hysteretic IVCs are commonly observed in granular superconducting films due to the Joule heating effect, which locally increases the effective temperature and hence reduces the $I_c$[38,39]. Further, at temperatures close to the $T_c$, intermediate resistive steps start to appear at 3.0 K and higher temperature as highlighted by the red dotted arrows in Fig. 3(b). The slopes of these resistive states increase in the increasing current direction and they converge at one single point [the grey dotted lines], known as the excess current $I_s$. These characteristics indicate that the intermediate resistive states originate due to PSLs[8,11]. Besides, the intermediate resistive steps become more prominent for the sample TN8 as shown in Fig. 3(c). A single sharp switching is observed at 2.4 K and below but the sudden emergence of intermediate resistive steps takes place for the isotherms measured at 2.5 K - 2.8 K. Further, black dotted lines connecting the slopes of each resistive steps meet at a single point on the current axis at $I_s$ confirming the formation of PSLs in disordered TiN thin films. These intermediate resistive steps get more prominent even at lower temperature for the sample TN12 as shown in Fig. 3(d). Here, the resistive states exist for much wider extent in current indicating the rduced $T_a$ indeed play a favorable role in support of PSL formation.

Reduction in thickness introduces more disorder[14,40-42] as that is also evident in the increased $R_n$ presented in Fig. 1(b) and Fig. 2(d). However, relating disorder directly to the resistivity would not be appropriate in the present scenario as there are secondary phases present along with TiN. Here, the disorder is related mainly to the granularity and the grain size of the films as the surface morphology studied by SEM and AFM [ Fig. S1-S3 in the SI] suggests the variations in the grain size with sample thickness. Generally, granular superconductors can be considered as an array of Josephson junctions where superconducting grains act as the superconducting islands and the grain boundaries serve as the weak -links and the macroscopic superconductivity is established by superconducting proximity effect. However, the establishment of macroscopic coherence depends strongly on the grain size and the inter-granular distance. Further, the intergranular region is very important as it contributes to the disorder by possessing any impurity like the presence of Si and the silicide phases, oxygen content, Ti residual layer etc. Here, for the thinner samples with much smaller grain size, the effect of grain boundaries, intergranular distance, the content and/or type of the intergranular region contribute significantly to the disorder as compared to that of the thicker samples. Furthermore, the substrate roughness [Fig. S4 in the SI] also becomes important for the thinner samples as that can induce disorder in the form of strain or probable weak links leading to PS.

In order to examine the effect of disorder on IVCs, we have studied more samples with reduced thickness and varying $T_a$. In Fig. 4, we present IVCs for two such samples TN5 and TN9 that are annealed at 820 °C & 780 °C, respectively. The samples from the 3 nm thickness batch. Here for sample TN5 with $T_a$ = 820 °C in Fig. 4(a), the intermediate resistive steps during the SC-to-NM transition become much more prominent and wide compared to that for 4 nm thick sample TN4 presented in Fig. 3(b). Initially, IVC isotherms show almost single step transition for measurement temperature 2.3 K and below, whereas, for temperature above 2.3 K, the single step transition is converted into a multiple step transition from superconducting state to metallic state. Furthermore, compared to TN5, intermediate resistive steps get even more pronounced and broad for sample TN9, which was annealed at 780 °C as shown in the Fig. 4(b). Unlike the sample TN8 annealed at 780 °C with thickness 4 nm presented in Fig. 3(c), wide resistive states start to show up even at 2 K base temperature for TN9. These resistive steps are a signature of the formation of PSLs occurring due to the annihilation of vortex-antivortex pairs emerging from the edges of the films[37,43]. Therefore, both $T_a$ & film thickness play crucial roles in altering the emergence of intermediate resistive steps in IVCs, however, the effect of film thickness is more prominent than that of $T_a$ as the former introduces more disorder than the later.

In order to examine the effect of magnetic field on the IVCs of these disordered TiN thin film samples, we have selected two samples, TN4 and TN5, that were annealed at $T_a$=820 °C but of thickness 4 nm and 3 nm, respectively. Here, IVCs were measured at varying magnetic field while keeping the measurement temperature unaltered. For sample TN4, the field dependent IVCs measured at 2.7 K are presented in Fig. 5(a) which shows single step transition from SC-to-NM state from 0 - 25 mT. With a further increase in the field up to 50 mT, an intermediate resistive step starts to appear in the course of the SC-to-NM transition instead of a single step transition. Here, the resistive step gets narrower in extent with increasing field and the same remains prominent up to 250 mT. With further increasing field, the intermediate step starts to diminish and the SC-NM transition gets smoother. Furthermore, the evolution of the intermediate resistive step with magnetic field has been displayed by the contour color plot in Fig. 5(b) with the support of the differential resistance (*dV/dI*) extracted from the measured field dependent IVCs. Considering the range of magnetic field within which the intermediate step appears, the contour plot in Fig. 5(b) is confined within 300 mT. Here, because of the single step transition, the differential resistance is maximum at 0 – 25 mT, whereas at higher fields, it gets reduced by the introduction of intermediate resistive steps that are indicated by the presence of multiple peaks at a particular field during the SC-NM transition. However, at 250mT and above, presence of a single, and relatively stronger peak, in the *dV/dI* indicates the removal of intermediate steps by the external magnetic field. In Fig. 5(b), the position corresponding to the peak of *dV/dI* at a particular magnetic field relates to the $I_c$ which reduces as the field moves towards the higher side. As observed in Fig. 4, reduction in film thickness makes the intermediate resistive steps more prominent. In order to examine the effect of field on reduced film thickness, we have measured field dependent IVCs for the sample TN5 with thickness 3 nm at the measurement temperature 2.3 K and the same is exhibited in Fig. 5(c). Here, the zero field IVC shows no intermediate steps but the steps start to emerge at 25 mT and remain prominent up to 175 mT. It is to be noted that with reduction in thickness from 4 nm to 3 nm, the number of intermediate steps increases indicating a greater number of transition segments in the SC-NM phase transition region. The same is evident by the appearance of multiple peaks in the concerned differential resistance *dV/dI* representation through the colour plot in Fig. 5(d). From Fig. 5(c-d), initially the intermediate steps get wider with increasing field up to about 100 mT and with further increasing field in the range between 125-175 mT, the wide steps start to break up into a greater number of steps. The splitting of the steps is clearly evident by the presence of multiple low-amplitude *dV/dI* peaks encircled by the dotted region in Fig. 5(d). Furthermore, intermediate sharp resistive steps start to disappear at 200 mT and there is a gradual increase in voltage from superconducting to normal state with a very little variation in differential resistance.

Further, we have extracted the critical current ($I_c$) and retrapping current ($I_r$) from the IVCs at zero field for five representative samples TN4, TN5, TN8, TN9 and TN12. Already, we have presented the effect of $T_a$ and film thickness on the IVCs of disordered TiN thin films in Fig. 3 & 4. Here, we show the influence of $T_a$ and film thickness on the temperature dependent $I_c$ & $I_r$. We have investigated the temperature dependence of $I_c$ in the light of Ginzburg- Landau (GL) theory[32] and modified GL theory[44] whereas, the $I_r$ has been explained by the existing hotspot model given by Skocpol, Beasley, and Tinkham (SBT)[38]. We have found that the theoretical GL equation $I_c \propto (1-t)^{s'}$; with $s' = 3/2$ is valid near the $T_c$ to explain the temperature dependent $I_c$ and it deviates at temperature far below the $T_c$ ($T < 0.75T_c$) as shown by the blue dotted curves in Fig. 6(a), (b) and (d) for samples TN4, TN8 and TN5, respectively. In order to fit entire range of measured temperature points, we have adopted the modified GL equation[44], $I_c \propto$ $(1-t^2)^s(1+t^2)^{1/2}$, where we use the exponent '$s$' as the free parameter to have the best fit to the experimental data. The red solid curves represent the best fits using the aforementioned modified GL equation for the samples presented in Fig. 6(a), (b) and (d) with the exponent values 1.32, 1.35 and 1.70, respectively. The acquired exponent ($s$) values from the fitting are close to the theoretical value of exponent around 1.5[32]. However, the difference of about 0.03 in the exponent $s$, for the samples with same thickness of 4 nm (TN4 & TN8), is negligible compared to the difference of about 0.35 for the sample TN5 of 3 nm thickness with the samples having thickness of 4 nm (TN4 or TN8). Further, for samples TN12 [Fig. 6(c)] and TN9 [Fig. 6(e)], the measured temperature varies within $0.75T_c$ and $T_c$, and we obtain the best fits (the solid blue curves) using the GL equation. where the exponent $s'$ is used as a free parameter. Here, the values for the exponent $s'$ from the fits are 1.31 and 1.83 for TN12 and TN9, respectively. The exponent for the sample TN12 of 4 nm thickness is close to the theoretical value[32] and also is similar to the other 4 nm thick samples TN4 &TN8. However, the deviation in the exponent for the sample TN9 with thickness of 3 nm from the theoretical value is more and is on the higher side. Furthermore, the fits from two other 3nm samples, TN5 & TN9, thickness have offered exponent values that are very close to each other. Therefore, it is evident from the above critical current analysis that the thickness, and hence, the disorder plays a significant role in controlling the temperature dependent $I_c$ in wide disordered superconducting thin films. Here, we observed that the exponent values, as obtained from the fits for 4 nm thick samples (TN4, TN8, TN12) are of the same order and lower than the theoretical value and 3 nm thick samples (TN5&TN9) have exponent values much higher than the theoretical value of 1.5. The higher exponent is mainly due to the presence of higher level of disorder in thinner samples. Apart from $I_c$, we have measured the temperature dependent $I_r$ for all the above five samples and the collection of those is presented in Fig. 6(f). First, we observe that the amplitude of $I_r$ reduces with the reduction in $T_a$ for the samples having same thickness. Similarly, reduction in thickness also reduces the $I_r$ for samples annealed at same $T_a$. Therefore, $I_c$ and $I_r$ follow the similar trend as that of $T_c$ with the variations in thickness and $T_a$. In the present case due to hysteretic IVCs, $I_c$ and $I_r$ are different and this is mainly due to the Joule heating in the resistive state[38,39]. Therefore, we have used the SBT hotspot model using equation, $I_r \propto (1-t)^{s''}$, to fit the retrapping current. Here, $s''=1/2$ is the theoretical value. The best fitted colored curves, using the exponent $s''$ as a free parameter, are presented in Fig. 6(f) for individual samples. The exponent for the samples TN4 ($s''=0.46$), TN8 ($s''=0.44$) & TN12 ($s''=0.42$) are close to the theoretical value of 0.5, whereas, there is a small deviation from the theoretical value for the sample TN5 ($s''=0.37$) and a further deviation appears for TN9 ($s''=0.23$). The deviation in exponent values from the theoretical value increases with decreasing $T_a$ and reduction in thickness indicating the influence of disorder on the mechanism of depairing current limit. Here depairing current refers to the maximum dissipationless current carried by a superconductor. Finally, the influence of $T_a$ and film thickness on the $I_c$ and $I_r$ is reflected by the various exponent values used to explain and understand the experimental results by using the existing models[32,38,44]. From the temperature dependent critical current analysis, the effect of $T_a$ on the exponent remains almost invariant ($s$ ~1.3), whereas, the effect of thickness plays a significant role in tuning the exponent values that shoot up with decreasing film thickness. Interestingly, the resistive steps appearing in the IVCs presented in Fig. 3 & 4, get more prominent and wider for reduced thickness rather than $T_a$. Finally, the reduced film thickness leads to suppression of $T_c$ & $I_c$, the appearance of PSLs in IVCs, and the deviation of the temperature dependent critical current from its theoretical value. Therefore, the enhanced level of disorder due to the reduction in film thickness plays a very important role in monitoring the superconducting properties and the overall SC-NM phase transition.

## Conclusion

In conclusion, we have fabricated disordered superconducting TiN thin films by establishing an unconventional but robust, nitridation method which involves high temperature annealing of Ti thin films deposited on $Si_3N_4$/Si substrates. The annealing process leads to the decomposition of $Si_3N_4$ into N and Si atoms that form majority TiN and minority $TiSi_2$ phases by reacting with Ti[18]. Generally, disorder is present in pure thin films and here the presence of $TiSi_2$ and excess Si can further enhance the level of disorder. The best obtained $T_c$ among the measured samples is ~ 4.8 K for film thickness of about 12 nm and above. As expected, $T_a$ plays a very important role in tuning the $T_c$. We also show that the $T_c$ and $R_n$ depend strongly on the film thickness. Furthermore, the IVCs for ultrathin (≤ 4 nm) samples feature staircase-like steps indicating the appearance of PSLs that evolve more prominently with the reduction in thickness and by external magnetic fields. Finally, the $T_c$ achieved by using this technique is comparable to the best reported values in the literature (Table S1). Other superconducting parameters are summarized in Table 1. Furthermore, the $T_c$, PSLs and the temperature dependent critical current vary in a similar fashion on the growth parameters like $T_a$ and the thickness.

## Methods

### *Material synthesis through a novel technique*

We used undoped intrinsic Si (100) substrate covered with 80 nm $Si_3N_4$ dielectric spacer layer, grown by low pressure chemical vapor deposition (LPCVD), as the substrate for the growth of TiN films. The substrates were cleaned by using the standard cleaning process involving sonication in acetone and isopropanol bath for 15 minutes each, followed by oxygen plasma cleaning of 10 minutes. Thereafter cleaned samples were loaded into the sputtering unit and transferred to ultra-high vacuum (UHV) chamber where they were pre-heated at about 820 °C for 30 minutes under very high vacuum conditions ( $p \sim 5 \times 10^{-8}$ Torr) in order to remove any adsorbed or trapped molecules/residue on the surface of the substrate. After performing all the cleaning process, cleaned $Si_3N_4$/Si substrate was cooled down to room temperature and transferred to the sputtering chamber through load lock without breaking the vacuum. A thin layer of Ti was then deposited on the substrate by dc magnetron sputtering of Ti (99.995% purity) in the presence of high purity Ar (99.9999%) gas. Sputtering of Ti was performed with a base pressure less than $1.5 \times 10^{-7}$ Torr. Finally, the sputtered sample was transferred again into the UHV chamber for post-sputtering annealing. Ti films on $Si_3N_4$/Si (100) were annealed at different annealing temperatures (~820 °C, ~780 °C, ~750 °C, ~700 °C and ~650 °C) for 2 hours at a pressure less than $5 \times 10^{-8}$ Torr. The thickness values for the films are determined by the optimized rate and thickness measurements on various samples prepared with same deposition time. The average value obtained from the thickness measurements is assigned to the thickness and the errors represent the range of variations on the measured values. Here, variation in annealing temperature is within ±10°C. During the annealing process, $Si_3N_4$ decomposes into Si and N atoms and due to high affinity of titanium towards both nitrogen and silicon, stable stoichiometric cubic titanium nitride (TiN) forms as a majority phase along with the minority phases such a titanium silicide ($TiSi_2$) and elemental silicon (Si). Here, temperature plays an important role to control the chemical kinetics happening during the chemical reaction because phenomena like decomposition and diffusion are very much sensitive to temperature. Therefore, concentration of majority and minority phases varies with the annealing temperatures ($T_a$) such as for $T_a > 780$ °C, majority phase is accompanied by elemental Si rather than silicide phase, whereas for $T_a \leq 780$ °C, it is accompanied by silicide phase. The concentration of silicide phase is maximum for $T_a \sim 780$ °C and the same gets reduced with the decrease in $T_a$.

For structural characterization we fabricated thin films on substrates with an area of 5 mm x 5 mm. For the transport measurements, we have patterned the thin films into Hall bar geometry by using shadow mask made of stainless steel. We have used a complimentary separate mask to make the contact leads for voltage and currents probes. The device geometry for a representative device is shown in the inset of Fig. 1 (a). The contact leads were made of Au (~50 nm)/Ti (~5 nm) deposited by dc magnetron sputtering.

***Structural characterizations***

Photoelectron Spectroscopic (XPS) measurements were performed in UHV based Multiprobe Surface Analysis System (Thermo Scientific Nexsa) operating at a base pressure of $1.5\times10^{-7}$ Pa. A monochromatic Al-Kα radiation source (1486.7 eV) was employed with radiation spot size of 400 μm for the excitation. To enhance the sampling depth and minimize the contribution of native oxides & contaminants, in-situ sputtering via energetic $Ar^+$ ions (at 500 eV) was performed for about 2 minutes. The depth profile scan was taken in “snapshot mode” in order to reduce the data acquisition time at the expense of spectra resolution. The adventitious carbon C-C bond at 284.8 eV was used for binding energy correction.

***Morphological Characterizations:***

The morphological characterizations of the samples were performed by using scanning electron microscope (Gemini SEM 500 ZEISS) and atomic force microscope (AFM) from BRUKER, Dimension ICON, in tapping mode.

***Low temperature transport measurements***

Transport measurements were carried out using a 16T/2K Physical Properties Measurement System (PPMS) of Quantum Design, USA at UGC-DAE CSR Indore.

**Data availability**

The data that represent the results in this paper and the data that support the findings of this study are available from the corresponding author upon reasonable request.

**Acknowledgements**

The authors highly acknowledge UGC-DAE CSR, Indore, India for carrying out the low temperature measurements in PPMS. We are thankful to the central facilities at IIT, Mandi, India for carrying out the XPS, SEM and AFM characterizations. We highly acknowledge Mr. Kumar Palit from IIT, Mandi, for his prompt response in carrying out the SEM and AFM characterizations within a very short period of time. We express our sincere thanks to Dr. Sudhir Husale for his support in SEM characterization. We are thankful to Mr. M. B. Chhetri, Mr. Bikash Gajar & Ms. Deepika Sawle for their assistance in the lab. S.Y. acknowledges the Senior Research fellowship (SRF) from UGC. Authors acknowledge the financial support for establishing the dilution refrigerator facility at CSIR-NPL from the Department of Science and Technology (DST), Govt. of India, under the project, SR/52/pu-0003/2010(G). This work was supported by CSIR network project 'AQuaRIUS' (Project No. PSC 0110) and is carried out under the mission mode project "Quantum Current Metrology".

**Author contributions**

S.Y. and S.S. fabricated the devices. S.Y., V.K., M.P.S., V.G. and R.P.A. carried out the low temperature transport measurements. S.Y. and S.S. analyzed the data and S.Y. wrote the manuscript. S.S. planned, designed and supervised the project. All the authors read and reviewed the manuscript.

**Additional information**

**Supplementary Information**

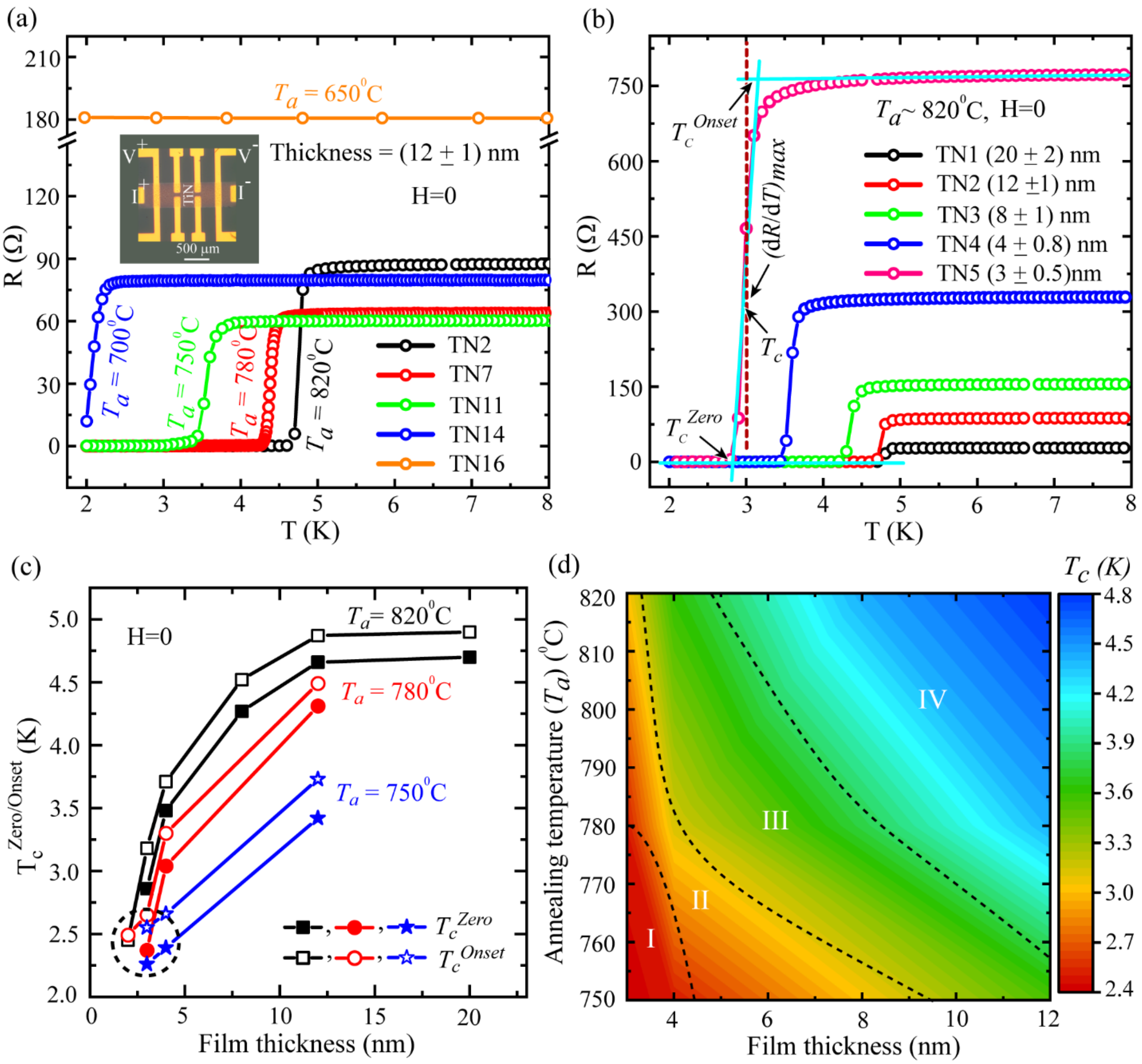

**Fig. 1:** Superconducting properties probed by temperature dependent resistance *R(T)* measurements. (a) A set of *R*(*T*) measurements done on TiN thin films grown at varying annealing temperature ($T_a$) while keeping the thickness same of about 12 nm. Inset: Optical microscopy image of a representative multiterminal device showing the measurement scheme for tranport measurements in four probe geometry. (b) A different set of *R(T)* measured on samples with varying thickness but annealed at same temperature, $T_a$ = 820°C. Different critical temperatures, ($T_c^{Onset}$, $T_c^{Zero}$ and $T_c$), are defined as shown in (b) for the representative curve with thickness around 3 nm. (c) Thickness dependent $T_c$ values collected from the samples annealed at $T_a$ = 820°C, 780°C and 750°C. (d) A contour color plot representation of the summary related to the dependence of critical temperature ($T_c$) on thickness and annealing temperature ($T_a$). Different regions are marked by Roman numerals and are bounded by the dashed boundary lines.

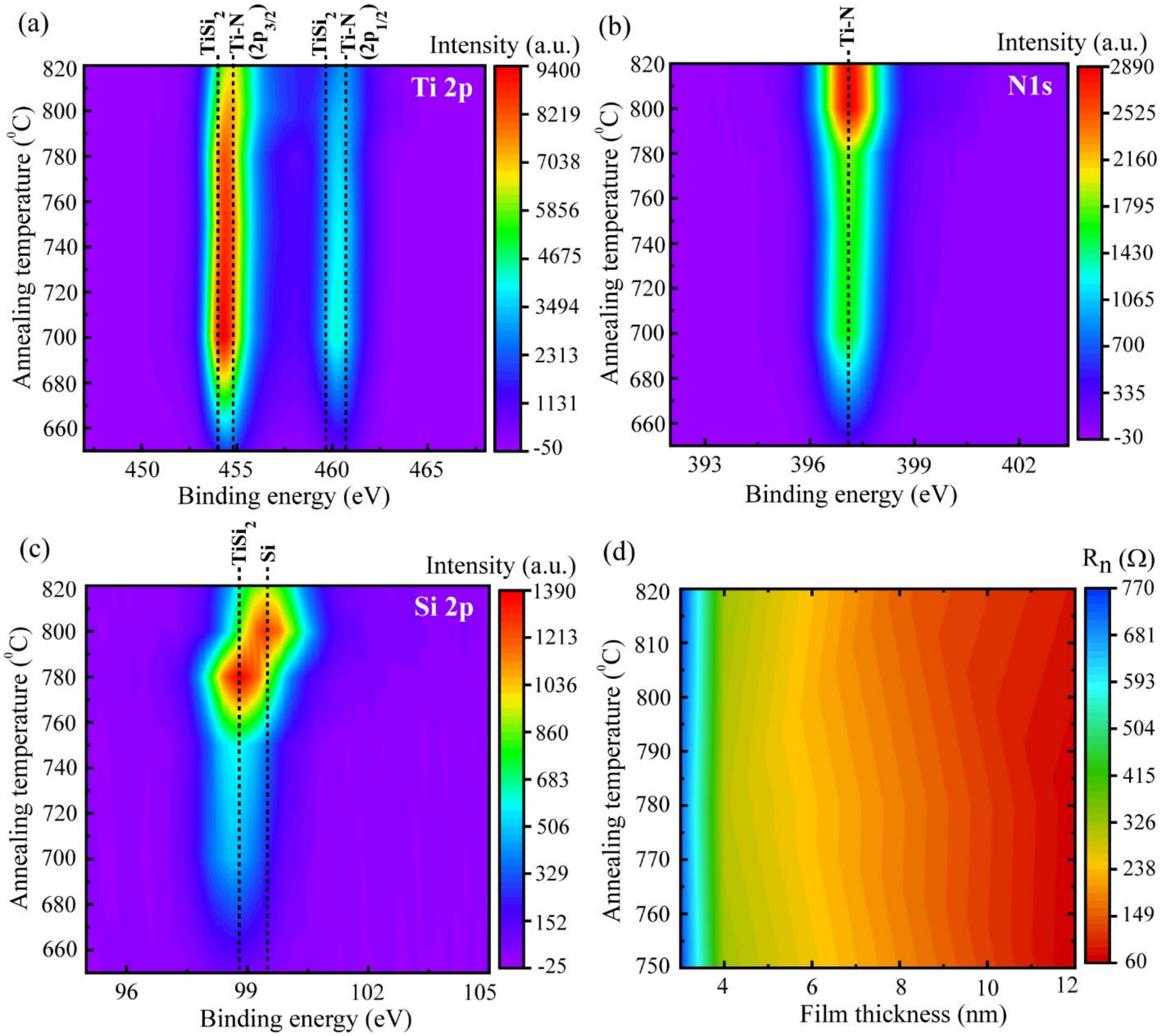

**Fig. 2:** Core-level XPS binding energy spectra of (a) Ti 2p, (b) N 1s and (c) Si 2p for reference samples annealed with varying annealing temperatures. The XPS spectra were taken from inside the films after removing the top surface by $Ar^+$ ion sputtering at 500 eV for about 2 minutes. (d) Variation in the $R_n$ with the variations in $T_a$ and thickness of the films.

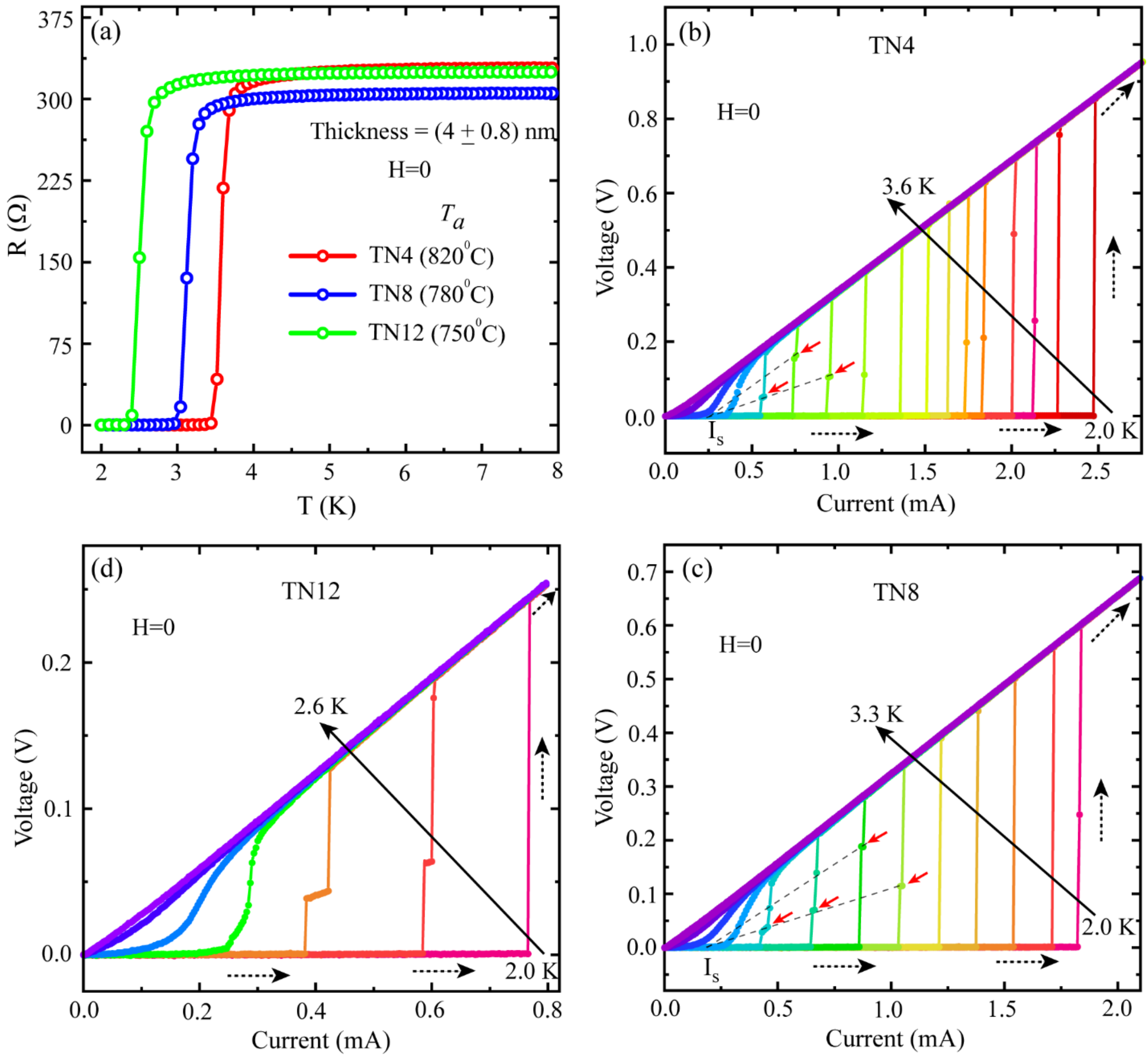

**Fig. 3:** Zero-field current-voltage characteristics (IVCs). (a) Collection of *R(T)* data measured on three selected samples for which the IVCs are displayed in (b-d). The three representative samples (TN4, TN8 & TN12) are of same thickness (~4 nm) but were grown at different annealing temperatures. The IVC isotherms for (b) TN4 ($T_a$ = 820°C), (c) TN8 ($T_a$ = 780°C) and (d) TN12 ($T_a$ = 750°C). The black dashed arrows indicate the sweeping direction of the bias current along with the transition from superconducting to normal state. The solid black arrows specify the extent in measurement temperature starting from the base temperature (2.0 K) up to the $T_c$ with 0.1 K interval. Here, only the upward direction of current sweeping is considered while presenting the IVCs. The red arrows show the appearance of intermediate resistive states.

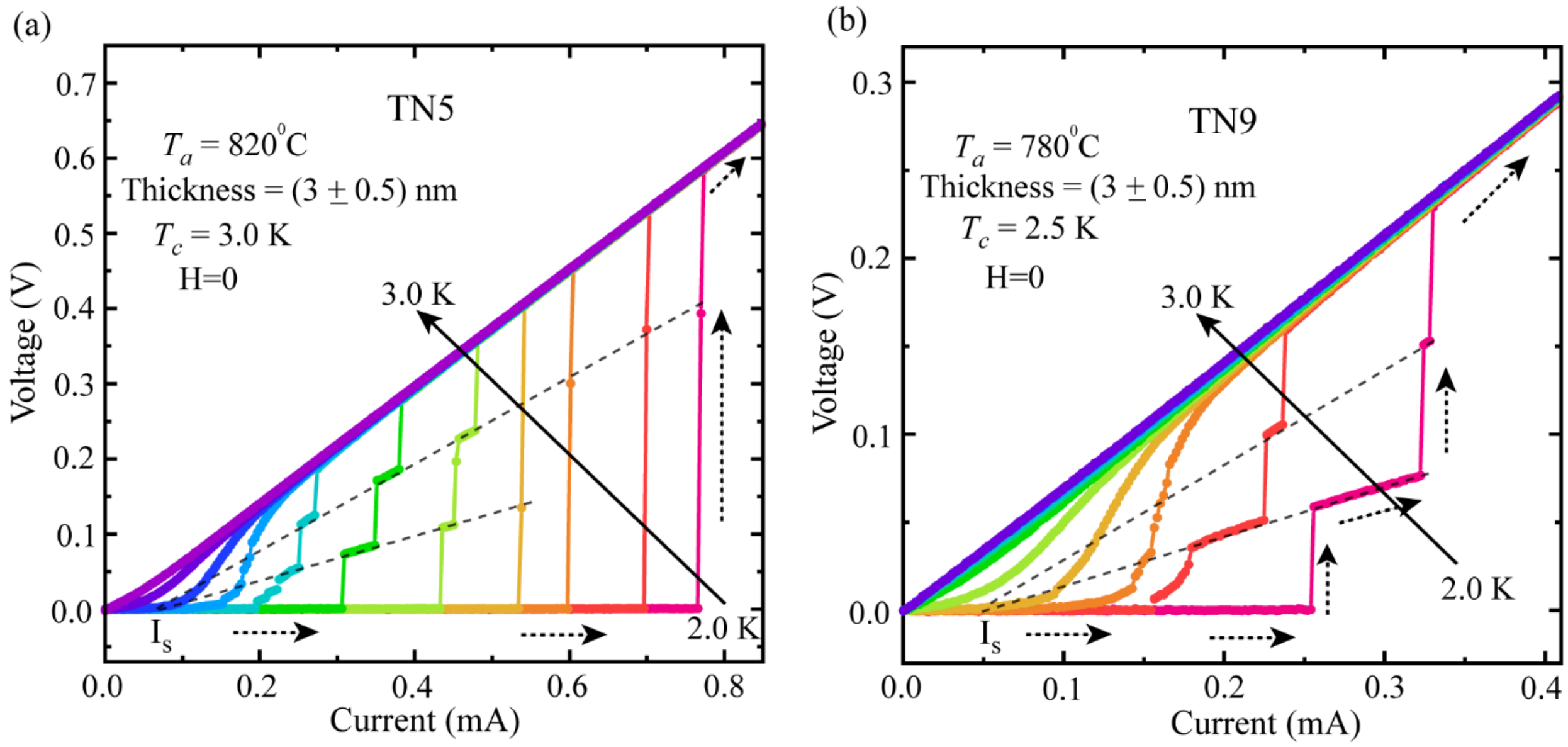

**Fig. 4:** Isothermal IVCs measured on samples with reduced thickness. Zero-field IVC isotherms for (a) TN5 (~3 nm) annealed at 820°C and (b) TN9 (~3 nm) annealed at 780°C. Here, the samples TN5 and TN9 are of same thickness of about 3 nm. The dotted black arrows indicate the sweeping direction. The grey dashed lines indicate the convergence of the resistive states at the excess current $I_s$.

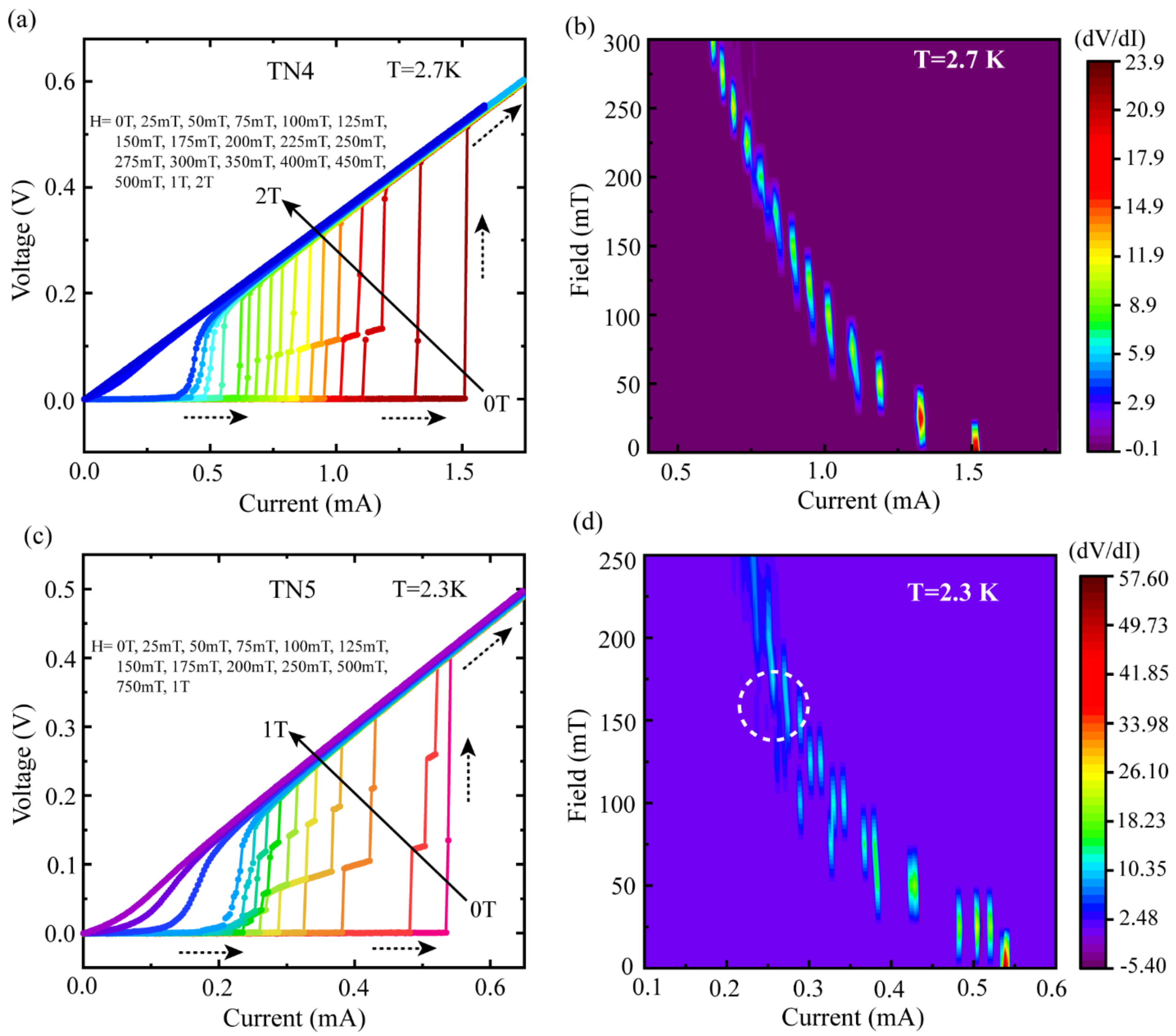

**Fig. 5:** Magnetic field dependence of IVCs for samples TN4 (~4 nm) and TN5 (~3 nm) that are annealed at 820°C. (a) Field-dependent IVCs for sample TN4 measured at fixed temperature 2.7 K and (b) corresponding *dV/dI* with respect to field & current in 3-D colour plot. (c) Variation of IVCs with magnetic field but at fixed temperature 2.3 K for sample TN5 and (d) the related dV/dI with respect to field & current in 3-D representation using contour colour plot. The Black solid arrows and dotted arrows in (a) &(c) indicate the increasing direction of applied magnetic field and the current sweeping direction, respectively.

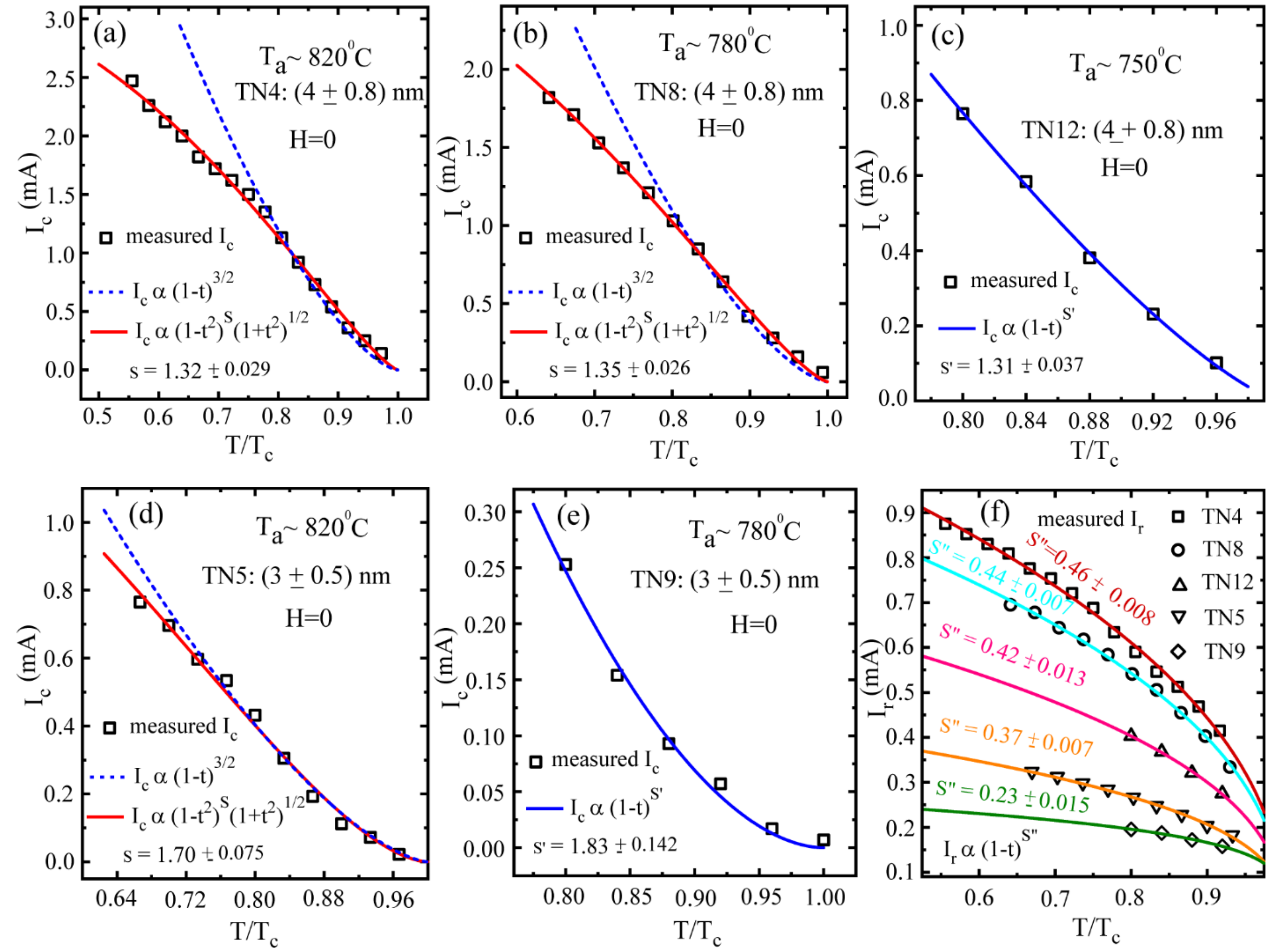

**Fig. 6:** Temperature dependence of characteristic currents such as critical current ($I_c$) and retrapping current ($I_r$). Dependence of $I_c$ on reduced temperature ($T/T_c$) for samples (a)TN4 annealed at 820°C, (b)TN8 annealed at 780°C, (c)TN12 annealed at 750°C, (d)TN5 annealed at 820°C and (e)TN9 annealed at 820°C. The samples in the first row, (a-c), are having the same film thickness of about 4 nm, whereas, the samples presented in (d) & (e) are having thickness of about 3 nm. The blue dotted lines indicate the fitting on the experimental data points close to the $T_c$ by using the Ginzburg- Landau (GL) eqn. $I_c \propto (1-t)^{3/2}$ , where t = $T/T_c$. The red solid lines are the fits obtained by using the modified GL eqn. $I_c \propto (1-t^2)^s(1+t^2)^{1/2}$, fitted over the entire range of measured temperature. However, blue solid lines in (c) & (e) give the best fit by using the GL equation with a free exponent s′ instead of fixing it to 3/2 for samples TN9 & TN12 at temperature close to the $T_c$. (f) $I_r$ related to all the five samples TN4, TN5, TN8, TN9 and TN12. For $I_r$, SBT model was used to fit the experimental data and the solid-colored lines represent the fitted curves by using eqn. $I_r \propto (1-t)^{s''}$

**Table 1: Superconducting parameters obtained for TiN thin films fabricated by using the substrate mediated nitridation technique**

| *Samples* | *$T_a$ , (°C) ± 10°C* | *d, (nm)* | *$T_C$, (K)* | *ΔT, (K)* | *ξ(0), (nm)* | *$R_n^{300K}$, (Ω)* | *RRR= ($R_n^{300K}$ / $R_n^{6K}$)* | *$\rho_n$ (6K), (μ-Ωcm)* | *Intermediate steps in IVCs @ zero-field* |
|---|---|---|---|---|---|---|---|---|---|
| TN1 | 820 | 20 ± 2 | 4.8 | 0.2 | ---- | 47 | 1.70 | 34.5 | ---- |
| TN2 | 820 | 12 ± 1 | 4.76 | 0.2 | 9.5 | 140 | 1.61 | 65 | ---- |
| TN3 | 820 | 8 ± 1 | 4.4 | 0.25 | ---- | 214 | 1.39 | 77 | ---- |
| TN4 | 820 | 4 ± 0.8 | 3.6 | 0.23 | 8.7 | 408 | 1.24 | 82 | Majorly single step transition |
| TN5 | 820 | 3 ± 0.5 | 3.0 | 0.32 | 7.94 | 879 | 1.14 | 135 | Multiple step transition |
| TN6 | 820 | 2 ± 0.5 | ---- | ---- | ---- | 25742 | 0.70 | 4588 | ---- |
| TN7 | 780 | 12 ± 1 | 4.38 | 0.18 | 10.8 | 98 | 1.54 | 47 | Single step transition |
| TN8 | 780 | 4 ± 0.8 | 3.12 | 0.26 | 9.2 | 370 | 1.21 | 75 | Multiple step transition |
| TN9 | 780 | 3 ± 0.5 | 2.5 | 0.28 | 9.0 | 808 | 1.10 | 127 | Multiple step transition |
| TN10 | 780 | 2 ± 0.5 | ---- | ---- | ---- | 2650 | 0.98 | 338 | ---- |
| TN11 | 750 | 12 ± 1 | 3.52 | 0.3 | 9.6 | 83 | 1.38 | 45 | Single step transition |
| TN12 | 750 | 4 ± 0.8 | 2.5 | 0.27 | 10 | 383 | 1.18 | 81 | Multiple step transition |
| TN13 | 750 | 3 ± 0.5 | 2.4 | 0.29 | ---- | 851 | 1.10 | 133 | ---- |
| TN14 | 700 | 12 ± 1 | 2.0 | 0.25 | ---- | 99 | 1.24 | 60 | ---- |
| TN15 | 700 | 4 ± 0.8 | ---- | ---- | ---- | 473 | 1.08 | 109 | ---- |
| TN16 | 650 | 12 ± 1 | ---- | ---- | ---- | 200 | 1.10 | 136 | ---- |

# Supplementary Information

# A Robust nitridation technique for fabrication of disordered superconducting TiN thin films featuring phase slip events

*Sachin Yadav[1, 2], Vinay Kaushik[3], M.P. Saravanan[3], R. P. Aloysius[1, 2], V. Ganesan[3] and Sangeeta Sahoo[1, 2,]**

[1]*Academy of Scientific and Innovative Research (AcSIR), AcSIR Headquarters CSIR-HRDC Campus, Ghaziabad, Uttar Pradesh, 201002, India.*

[2]*Electrical & Electronics Metrology Division, National Physical Laboratory, Council of Scientific and Industrial Research, Dr. K. S Krishnan Road, New Delhi-110012, India.*

[3]*Low Temperature Laboratory, UGC-DAE Consortium for Scientific Research, University Campus, Khandwa Road, Indore- 452001, India*

**Correspondences should be addressed to S. S. (Email: sahoos@nplindia.org)*

# Contents:

## 1. Scanning electron microscopy (SEM) images for selected samples

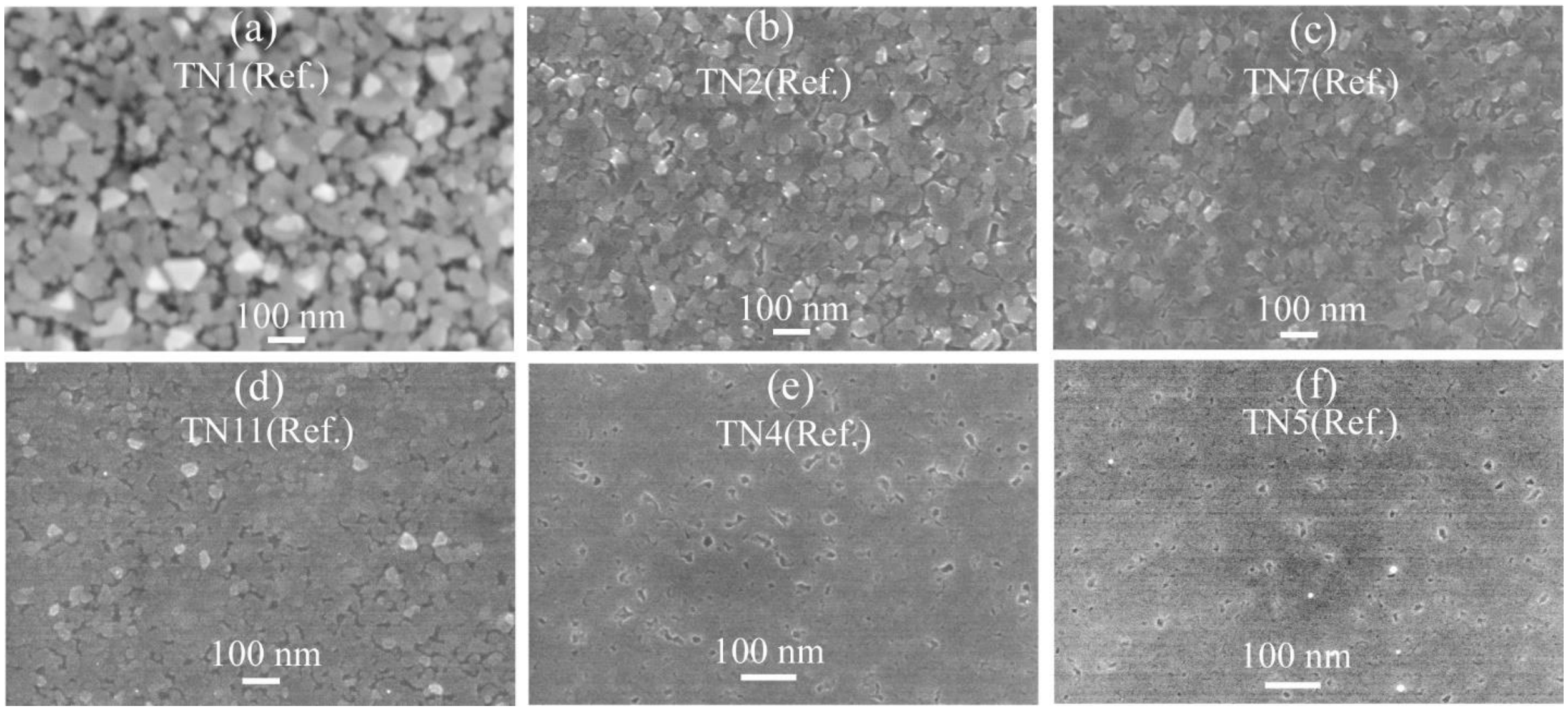

***Fig. S1:*** *Scanning electron microscopy (SEM) images for some of the selected devices. Here, the representative samples are the reference devices that were prepared at the same run/batch of the respective measured devices. The images represent the surface morphology of the samples and the granular nature is evident. The grain size strongly depends on the annealing temperature $T_a$ and the thickness.*

Scanning electron microscopy (SEM) imaging has been used to investigate the surface morphology of the measured devices and the corresponding SEM images are shown in Fig. S1. Here, we have selected identical devices that were prepared at the same run/batch of the respective measured devices and each reference sample was placed at the closest vicinity of the related measured device. For example, the sample TN1(Ref.) is the sample prepared along with the measured sample TN1 and the two samples were placed next to each other. The annealing temperature $T_a$ and the thickness values ($t$) for the samples are mentioned in Table 1 in the main manuscript. Here, the relatively thicker samples with varying $T_a$ are presented in Fig. S1(a)-(d), whereas, the two thinner samples with same $T_a$ have been presented in (e)-(f). The images suggest that the samples are granular in nature. Grains are bigger in size and they are clearly visible in the SEM images for the thicker sample. However, with changing thickness from 20 nm [Fig. S1(a)] to 12 nm [Fig. S1(b)], we observe that the surface roughness decreases and the surface looks smoother for the thinner sample. The surface gets further smoothed for reduced annealing temperature but with same thickness as it is clear in Fig. S1(b)-(e), for $T_a$ 820°C, 780°C and 750°C, respectively. The average grain size for these relatively thicker samples is about 40-50 nm. Here, the grains are clearly visible in SEM images for the thicker samples, TN1, TN2, TN7 and TN11. But the grains are not clear and there is a drastic change in the surface morphology for the samples TN4 & TN5 with thickness ~ 4 nm and 3 nm, respectively. Here, the thickness is playing a crucial role and a totally different surface morphology, containing very small grains that are not clearly resolved under SEM, is obtained. In order to resolve the grains for the thinner samples, the sample topography is further studied by using atomic force microscopy (AFM) imaging which is followed in the next section.

2. **Atomic force microscopy (AFM) images in 2D and 3D representation**

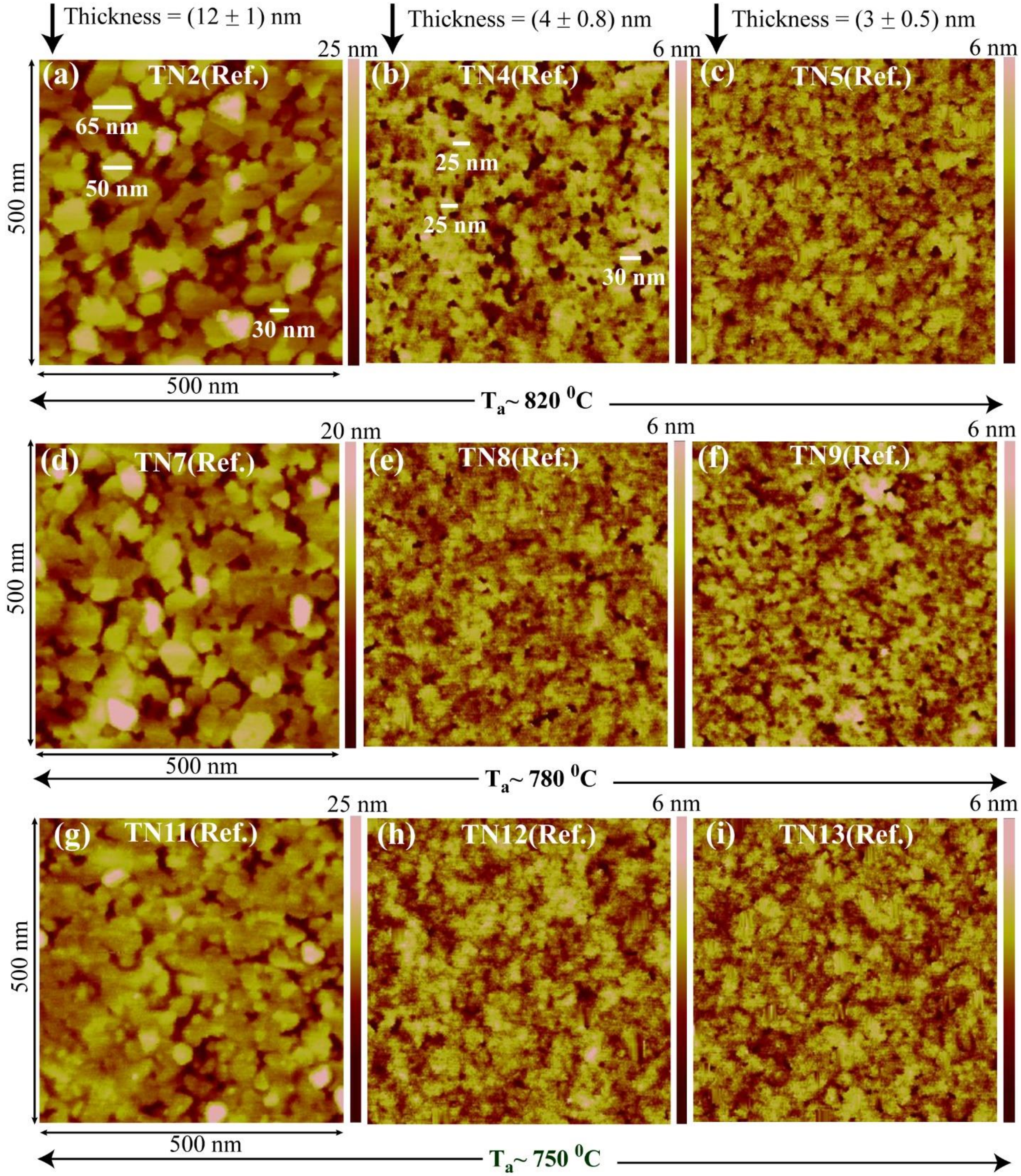

***Fig. S2:*** *Atomic Force microscopy (AFM) images for the reference samples corresponding to all the devices presenting I-V measurements in the main manuscript. Each row represents a particular $T_a$ and each column represents a particular thickness. $T_a$ decreases from top to bottom whereas, thickness decreases from left to right. For having an idea about the grain size, we have marked few grains for thicker and thinner samples in (a) and (b), respectively.*

In order to have better understanding about the surface morphology, surface roughness and grain size, we have carried out the topographic study by using atomic force microscopy (AFM) imaging. In Fig. S2, we have displayed a set of AFM images for all the samples that were characterized by IVCs in the main manuscript. The top row, middle row and the bottom row correspond to $T_a$ as 820°C, 780°C and 750°C, respectively. On the other hand, the first, second and third columns from the left represent the thickness category of 12 nm, 4 nm and 3 nm, respectively. Here, as already explained in the method section that the samples are categorized with a particular thickness which is determined by the optimized rate and from the thickness measurements obtained from various samples prepared with same deposition time. The error bars represent the range within which the measured thickness varied. As we have seen from the SEM images presented in Fig. S1, the thicker samples in first column of Fig. S2 display clear granular morphology with relatively large grain size. There is merely any difference in grain size for the samples TN2 ($T_a$ = 820°C) and TN7 ($T_a$ = 780°C), whereas, the surface looks smoother with smaller grain size for TN11 ($T_a$ = 750°C). For an estimate of the grain size variation, we have marked few grains with lateral dimension for TN2 in Fig. S2 (a). A large variation of about 30-65 nm in the grain size is observed for the sample in Fig. S2 (a). Here, it should be noted that the grain size for thicker samples is about 45-50 nm as estimated from the line width in X-ray diffraction pattern [Ref. #18 in the main manuscript: *Appl. Surf. Sci*. **541**, 148465 (2021)] which is in good agreement with the values obtained from the SEM images. Contrary to the thicker samples, the middle and right columns, presenting the thinner samples, show totally different type of surface topography containing patchy type of structures of about 25-30 nm in lateral dimension as shown by the marked patches in Fig S2(b). All the samples in these two columns look almost similar and only the surface coverage in the middle columns appears to be bit more than that of the right column. It should be noted that the thickness difference is about 1 nm between these two columns and obviously, it is very hard to find them different from each other.

Granular superconductors can be considered as an array of Josephson junctions where superconducting grains act as the superconducting islands and the grain boundaries serve as the weak -links and the macroscopic superconductivity is established by superconducting proximity effect (PE). By lowering temperature, the individual superconducting grains become superconducting locally when the temperature reaches to their transition temperature ($T_C$). Further the superconducting grains couple to each other by superconducting PE through the grain boundaries and the progressive coupling through PE establishes the global superconductivity in the granular matrix by establishing a macroscopic coherence. However, the establishment of macroscopic coherence depends strongly on the grain size for a particular material and the inter-granular distance. For example, when the superconducting grains are larger than the superconducting coherence length and they are closely spaced (the intergranular distance is much shorter than the respective coherence length), the global superconductivity is established within a short span of temperature and the metal-superconductor transition in $R(T)$ characteristics appears sharp. But, when the superconducting grains become comparable to characteristic length scales of electronic confinement and superconducting coherence length, finite size effects and superconducting fluctuation play a crucial role to control the metal-superconductor transition and the consequences are broad transition and incomplete superconductivity as the superconducting proximity is going to be impaired by the order parameter fluctuations. As the surface morphology, studied by SEM and AFM, suggests the variations in the grain size with sample thickness, the granularity and the grain size play a crucial role to modulate the transport properties.

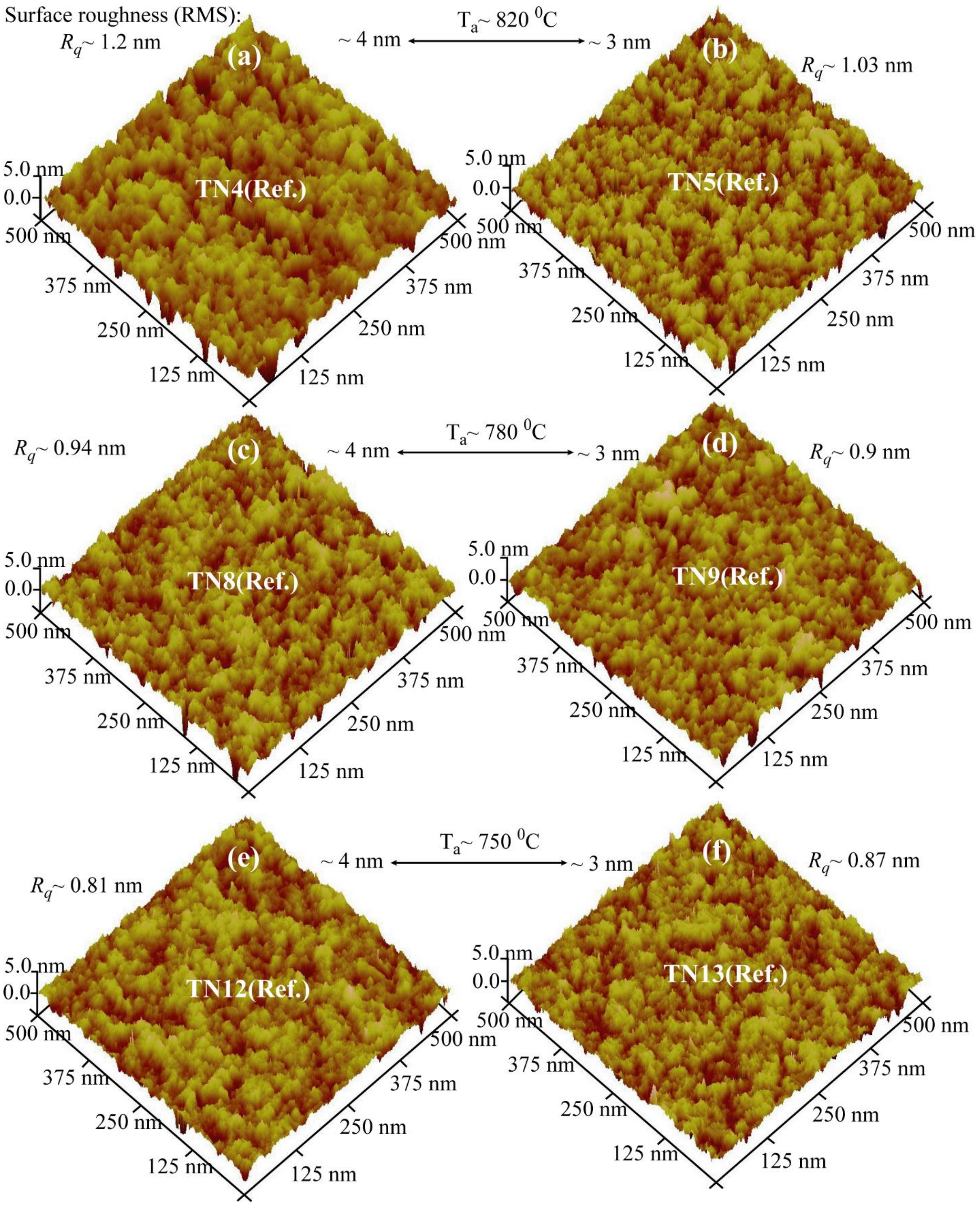

***Fig. S3:*** *AFM images in three-dimensional (3D) representation for the thinner samples that were shown in the middle and right columns in Fig S2.*

In order to understand the constituents of those patchy structures appearing in the thinner samples, we have displayed the AFM images in 3-dimensional (3D) representation in Fig. S3. The 3D AFM images clearly demonstrate that the patchy structures are consisting of fine grains joined together. These grains are very small in dimension and to resolve them in AFM is also very difficult. However, the granular nature along with their coverage is evident from Fig. S3. The surface roughness, which is the R.M.S roughness obtained from the images for each sample, is mentioned in the respective figures.

3. **Surface roughness from AFM images and its dependence on annealing temperature (*$T_a$*) and thickness**

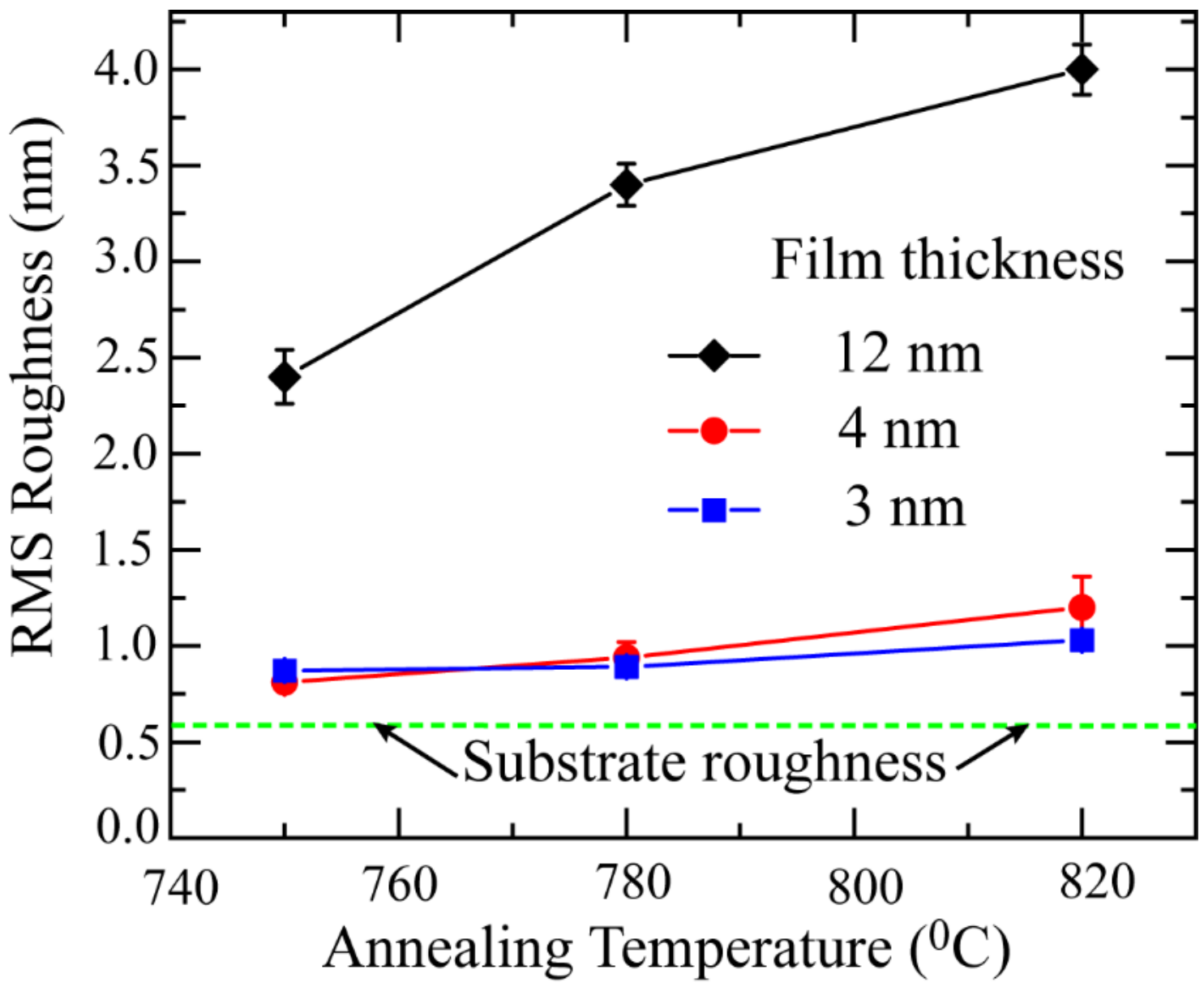

***Fig. S4:*** *Dependence of RMS surface roughness, obtained from the AFM images, on the annealing temperature (Ta) and on the film thickness. Here, the representative three thickness categories have been selected for three selected annealing temperatures, 820℃,780℃ and 750℃. The green dashed horizontal line represents the surface roughness for the substrate.*

In Fig. S4, we have plotted the RMS surface roughness values for all the samples with respect to the annealing temperature. Here, the roughness values are taken as the average of three AFM scans for individual samples and the error bars represent the range of the variation. First of all, we see that the roughness values are much higher for the samples of thickness category 12 nm. This was evident from both SEM and AFM images presented in Fig. S2 and Fig. S3. Further with increasing $T_a$, roughness increases which might explain the increment in the normal state resistance with increasing $T_a$ as presented in Fig. 1 in the main manuscript. The substrate roughness is shown by the green dashed horizontal line which is very close to the surface roughness values for the thinner samples as compared to that of the thicker ones. Hence, the effect of the substrate roughness is important for the thinner samples whereas that can be neglected for the thicker samples.

## 4. Calculation of Ginzburg- Landau (GL) coherence length ($\xi_{GL}$) for TiN samples prepared under different growth conditions.

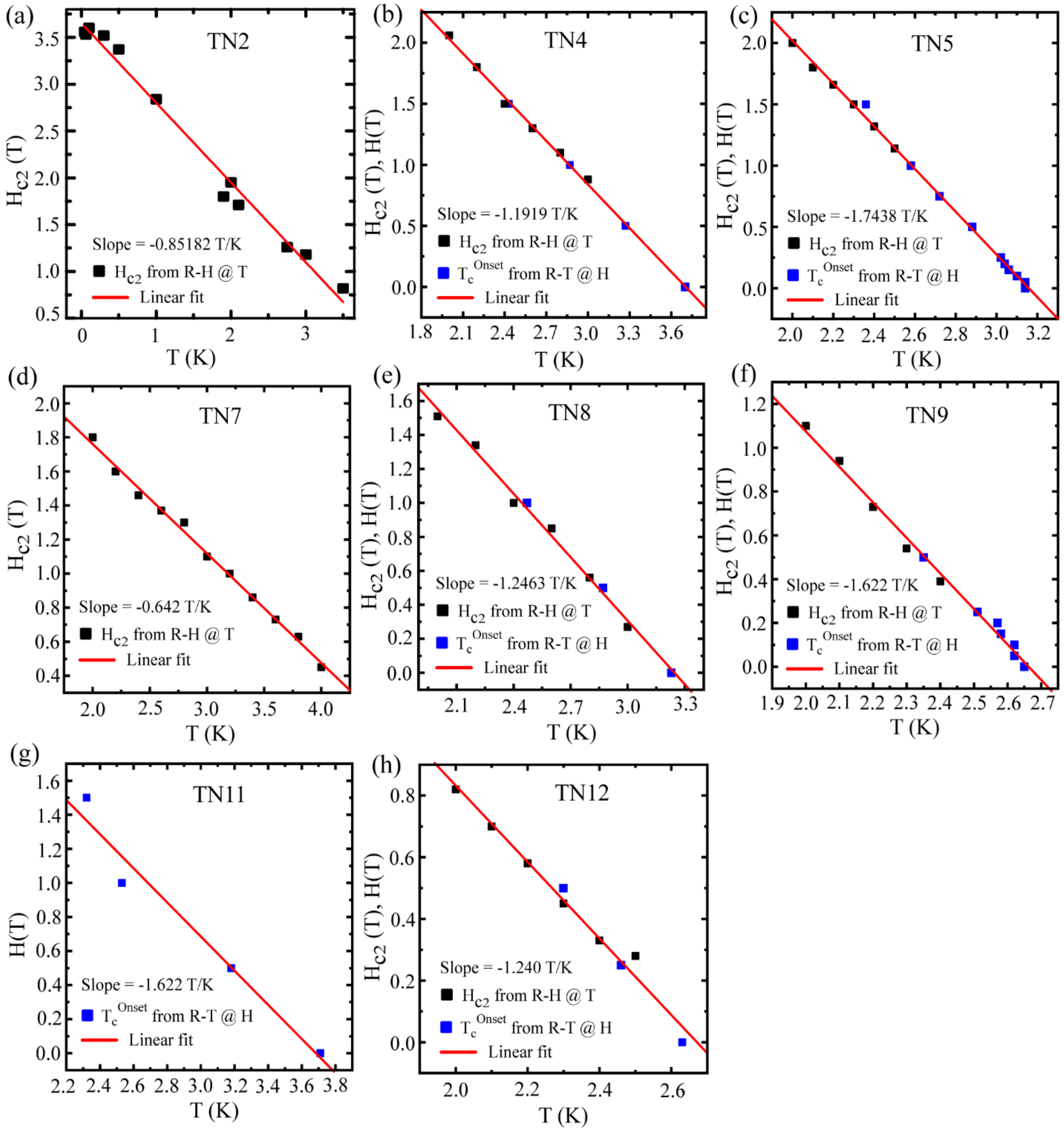

***Fig. S5:*** *B-T phase diagram for the TiN samples. Black and blue squares are the data points collected from isothermal magnetoresistance measurements [R-H@T] and temperature dependent resistance measurements under applied field [R-T@H], respectively. Solid red lines represent the linear fits performed on the experimental data points and provide the slope for calculating the GL coherence length* $\xi_{GL}$.

We have calculated the Ginzburg-Landau (GL) coherence length $\xi_{GL}(0)$, by using the standard formula, $\xi_{GL}(0) = \left[\frac{\phi_0}{2\pi T_c \left|\frac{dH_{c2}}{dT}\right|_{T_c}}\right]^{1/2}$, where $\phi_0$ is the flux quantum. The experimental data points for the samples TN2, TN7 &TN11 are the values of the upper critical field ($H_{c2}$) taken from magnetoresistance isotherms, whereas for sample TN11, experimental data points are the $T_c^{Onset}$ values taken from the field dependent *R(T).* However, for rest of the samples, the experimental data points are taken from both *R(T)* & *R(H)* as shown with the help of black & blue squares in Fig. S5. The extracted values from *R(T)* & *R(H)* are fitted linearly in Fig. S5 as shown by the red line. The slopes obtained from the linear fits were used for calculating the coherence length $\xi_{GL}(0)$ for all TiN samples fabricated under different growth conditions and the corresponding coherence lengths are 9.5 nm (TN2), 8.7 nm (TN4), 7.94 nm (TN5), 10.8 nm (TN7), 9.2 nm (TN8), 9.0 nm (TN9), 9.6 nm (TN11) & 10 nm (TN12), respectively.

## 5. IVCs of TiN samples showing both sweeps (up and down)

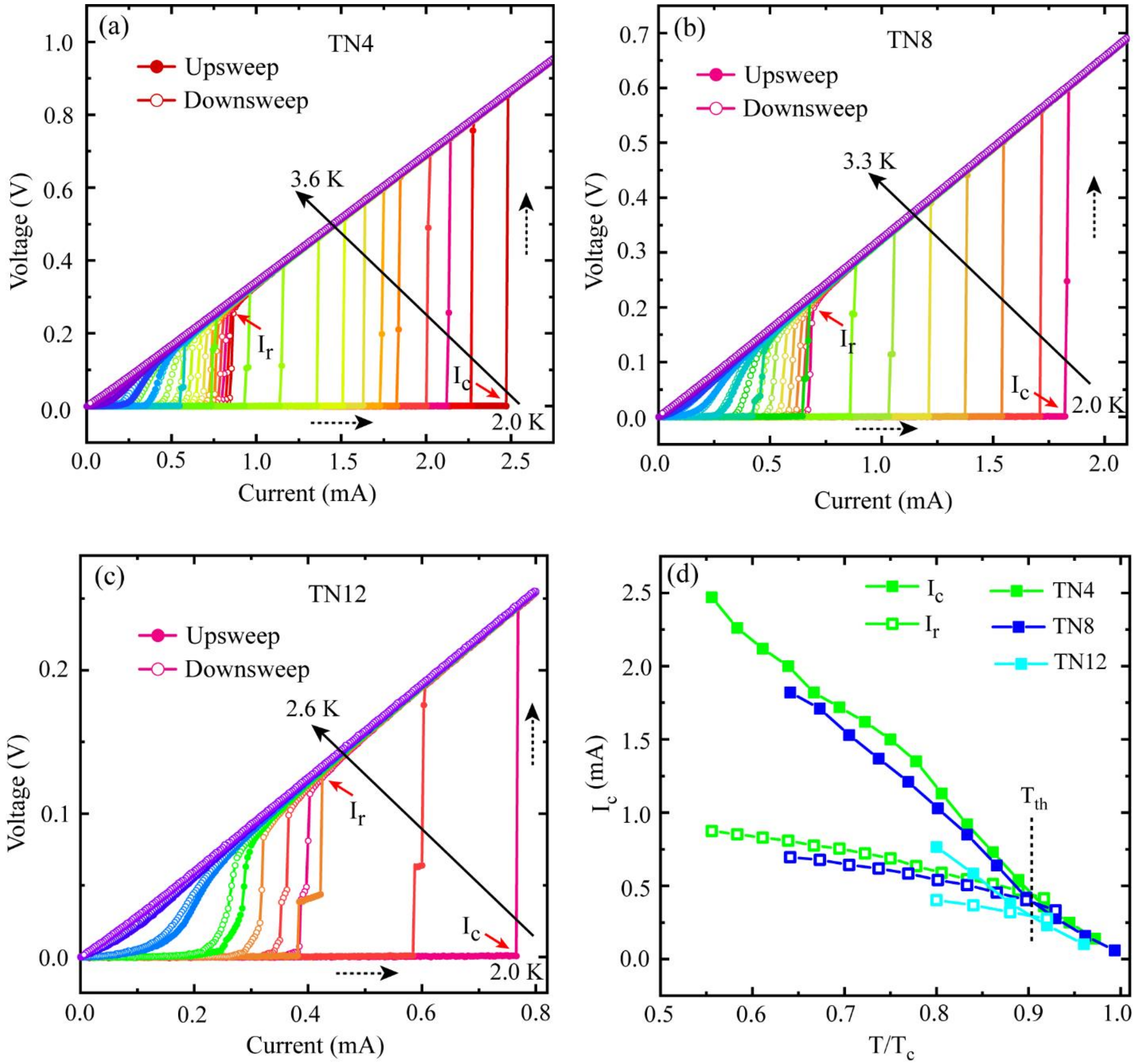

***Fig. S6:*** *Current voltage characteristics (IVCs) of TiN thin film samples showing up and down current sweeping directions. IVC isotherms for (a) sample TN4 annealed at 820 ℃, (b) sample TN8 annealed at 780 ℃ & (c) TN12 annealed at 750 ℃. (d) The dependence of critical current ($I_c$) & retrapping current ($I_r$) on reduced temperature ($T/T_c$). All the three samples are having the same film thickness which is about 4 nm. Solid and empty circles in the IVC isotherms represent the up and down current sweeping directions, respectively. Current sweeping direction for upsweep is shown by black dotted arrows.*

In Fig. S6, we have presented the current voltage characteristics (IVCs) of TiN thin film samples annealed at different growth temperature but having the same film thickness of about 4 nm. At lower temperature, the IVCs show single step sharp transition from superconducting to metallic state. The current at which the transition occurs is known as the critical current ($I_c$), whereas, the current corresponding to the onset of superconducting state from the normal state is defined as retrapping current ($I_r$). $I_c$ and $I_r$ are marked by the solid red arrows in Fig. S6(a). Moreover, we do see similar kind of single step transition in *IVCs* for sample TN8 as shown in Fig. S6(b). However, multiple resistive steps appear in the transition from superconducting to metallic state for the sample TN12 as shown in Fig. S6(c). Further, as these two characteristic currents ($I_c$ & $I_r$) are different, the IVCs appear to be hysteretic with respect to the current sweeping direction. With increasing temperature, the hysteresis gets reduced and vanishes at $T_C$. Hysteretic IVCs are commonly observed in granular superconducting films due to Joule heating effect, which locally increases the effective temperature and hence reduces the critical current. The dependence of $I_c$ and $I_r$ on temperature is shown in Fig. S6(d). Here, at lower temperature (far away from $T_c$), hysteresis is more prominent and it reduces as the temperature approaches close to the $T_c$. However, at a particular temperature, $I_r$ becomes equal to $I_c$ and the temperature is known as the threshold temperature ($T_{th}$) which is marked in Fig. S6(d) with a vertical dotted line and above this temperature $I_r$ becomes higher than $I_c$.

## 6. Table S1: Comparison of critical temperature ($T_c$) values for TiN reported in literature with the $T_c$ obtained in the present work

| *Reference | $T_C$ (K) | Thickness, (nm) | Growth technique | Substrate | Crystallographic Structure |
|---|---|---|---|---|---|
| Ref #21 | 4.6 | 80 | Reactive sputtering | Sapphire | |
| | 3.6 | 22 | | | |
| | 2.6 | 15 | | | |
| Ref #23 | 4.5 | 35 | Reactive sputtering | Si (100) | Poly-crystalline |
| Ref #24 | 0.7 K≤$T_C$≤4.5K | 20 nm≤ t ≤100 nm | Reactive sputtering | Si (100) | Poly-crystalline |
| Ref #25 | 4 K<$T_C$ <4.5 K ($T_C$ is determined by 90% resistance criteria) | 62 nm | DC biased sputtering | HF-cleaned Si (100) | Poly-crystalline; with dominant (200) orientation |
| Ref #29 | 6.0 | Size: 1x1x1mm$^3$ | Chemical-vapor Deposition | | Single crystal |
| Ref #30 | 5.25 | 40 nm | Plasma enhanced MBE | MgO (001) | |
| Ref #22 | 5.4 | ~ 200 nm | Reactive sputtering | HF-cleaned Si (100) | Single crystal |
| Ref #26 | 4.5 K | 100-200 nm | Reactive sputtering | HF-cleaned Si (001) | Poly-crystalline |
| Ref #27 | 3.0 K | 8.9 nm | ALD | Si (111) | Preferentially (200) oriented |
| | 4.6 K | 109 nm | | | |
| Ref #28 | 3.4 K | 18 nm | Plasma-enhanced ALD | Si (110) | |
| Ref #31 | 4.83 K | ~ 100 nm | PLD | MgO, SiN, Sapphire | Poly-crystalline |
| ***Current work*** | ***3.0 K≤$T_C$≤4.8 K*** | ***3 nm ≤ t ≤20 nm*** | ***Substrate mediated nitridation*** | ***SiN*** | ***Poly-crystalline*** |

* References are mentioned in the main manuscript

In Table S1, we have presented a list of references that represent the best achieved critical temperature ($T_c$) related to superconductor-metal phase transition for TiN. The last row of the table contains the results from the current work. Along with the $T_c$ values, we have also collected the film thickness, adopted techniques and the crystallographic structure. First of all, we find that our results are comparable and compared to many of the reported results, they offer even better $T_c$, particularly if we pay attention to the thickness range. However, higher $T_c$ values are reported in References 23, 25 and 24. The highest reported $T_C$ for this material is obtained from Ref #23 which actually dealt with bulk single crystal of size 1x1x1mm$^3$ prepared by CVD technique. It is obvious that the single crystalline materials are expected to offer a better $T_C$ compared to poly-crystalline films. In Ref # 23 & 24, the material is single crystalline. Further, the growth technique is very important for the quality of the materials. For example, MBE is known for high quality epitaxial thin film growth as appeared in Ref # 25 with higher $T_C$ for this material grown on MgO with nearly same lattice parameter.

For polycrystalline TiN films, usually grown by conventional reactive magnetron sputtering, the current work presents the most promising results as far as the superconducting critical temperature is considered. Here, we should note that all the reported works, presented in the table corresponding to $T_c$ values higher than 4 K, are obtained from much thicker films than the present work. Even though above a certain thickness, $T_c$ is supposed to be independent of thickness but we do observe variations in $T_c$ values with thickness (e.g. References 17, 29, 20 and in the current work) but in our case $T_c$ becomes independent of thickness at 15nm and above.